{
\documentclass[floats,showpacs,twocolumn,prb]{revtex4}%

\usepackage{amsfonts}
\usepackage{amsmath}
\usepackage{amssymb}
\usepackage[dvips]{color}
\usepackage{graphicx}%
\setcounter{MaxMatrixCols}{30}
\providecommand{\U}[1]{\protect\rule{.1in}{.1in}}
\providecommand{\U}[1]{\protect\rule{.1in}{.1in}}
\providecommand{\U}[1]{\protect\rule{.1in}{.1in}}

\ifx\pdfoutput\relax\let\pdfoutput=\undefined\fi
\newcount\msipdfoutput
\ifx\pdfoutput\undefined\else
\ifcase\pdfoutput\else
\msipdfoutput=1
\ifx\paperwidth\undefined\else
\ifdim\paperheight=0pt\relax\else\pdfpageheight\paperheight\fi
\ifdim\paperwidth=0pt\relax\else\pdfpagewidth\paperwidth\fi
\fi\fi\fi
\begin{document}
\title{Storage of Quantum Coherences as Phase Labeled Local Polarization in Solid
State NMR}
\author{Mar\'{\i}a Bel\'{e}n Franzoni}
\email{franzoni@mpip-mainz.mpg.de}
\affiliation{Facultad de Matem\'atica, Astronom\'\i a y F\'\i sica and Instituto de F\'\i
sica Enrique Gaviola (CONICET), Universidad Nacional de C\'ordoba, 5000 C\'ordoba, Argentina}
\author{Rodolfo H. Acosta}
\email{racosta@famaf.unc.edu.ar}
\affiliation{Facultad de Matem\'atica, Astronom\'\i a y F\'\i sica and Instituto de F\'\i
sica Enrique Gaviola (CONICET), Universidad Nacional de C\'ordoba, 5000 C\'ordoba, Argentina}
\author{Horacio M. Pastawski}
\email{horacio@famaf.unc.edu.ar}
\affiliation{Facultad de Matem\'atica, Astronom\'\i a y F\'\i sica and Instituto de F\'\i
sica Enrique Gaviola (CONICET), Universidad Nacional de C\'ordoba, 5000 C\'ordoba, Argentina}
\author{Patricia R. Levstein}
\email{patricia@famaf.unc.edu.ar}
\affiliation{Facultad de Matem\'atica, Astronom\'\i a y F\'\i sica and Instituto de F\'\i
sica Enrique Gaviola (CONICET), Universidad Nacional de C\'ordoba, 5000 C\'ordoba, Argentina}


\begin{abstract}
Nuclear spins are promising candidates for quantum information processing
because their good isolation from the environment precludes the rapid loss of
quantum coherence. Many strategies have been developed to further extend their
decoherence times. Some of them make use of decoupling techniques based on the Carr--Purcell and Carr--Purcell--Meiboom--Gill pulse sequences. In many cases, when applied to inhomogeneous samples, they yield a magnetization decay much slower than the Hahn echo. However, we have proved that these decays cannot be associated with longer decoherence times as coherences remain frozen. They result from coherences recovered after their storage as local polarization and thus they can be used as memories. We show here how this  freezing of the coherent state, which can subsequently be
recovered after times longer than the natural decoherence time of the system, can be generated in a controlled way with the use of field gradients. A similar behaviour of homogeneous
samples in inhomogeneous fields are demonstrated.
It is emphasized that the
effects of inhomogeneities in solid state NMR, independently of their origin, should not be disregarded as they play a crucial role in multipulse sequences. 

\end{abstract}
\startpage{1}
\maketitle


\section{Introduction}

The initial proposals for quantum information processing \cite{QCompgral}
triggered a worldwide effort to develop simple specific realizations.
\cite{Cory97,Bennett00}. In the last decade, much effort has been devoted to
understand and control decoherence \cite{zurek03} as it has become the main
drawback for quantum information processing
\cite{Zurek_Cucchietti_Paz_2006,alvarezJCP2006,DobrovitskiJPCM07,alvarezPRA2007,claudia09}%
. 

Within solid state Nuclear Magnetic Resonance the interest focuses in those
systems which have been proposed as possible candidates for quantum registers
~\cite{suter04,suter06,barrett,Ladd05}, quantum channels
~\cite{Cappellaro-CoryPRL2007,Cappellaro-Cory2007,elena09}, simulators of
quantum dynamics ~\cite{LUP98,alvarezPRL2010} or probes for specific quantum
gates \cite{alvarezJCP2006}.

More, recently, a number of interesting works seek to control the nuclear and
electron spin entanglement in systems where the nuclear spins can be used as
quantum spin memories such as N$@$C$_{60}$~\cite{mehring_nc60}, or $^{31}$P
~\cite{morton_solid-state_2008} or nitrogen vacancies (NV centers) in diamond
\cite{childress06,zhao_anomalous_2011,fuchs_quantum_2011}. Some of the
strategies implemented with NMR and EPR, such as decoupling techniques
\cite{CoryCC06,dobrovitskiSc_2010,bluhm_long_2010,cory_diamond_2010,barthel_interlaced_2010}%
., have a wide interest as they can be adapted to many other two level systems
like trapped ions, quantum dots\cite{Petta-Marcus2005} or Josephson
junctions~\cite{Martinis-Urbina1989,Vion-Urbina2002,hanson_spins_2007,biercuk_optimized_2009}%
. The renovated interest in pulse sequences that help to fight decoherence has
revived the old sequences such as the Hahn Echo \cite{Hahn}\textbf{,} the
Carr--Purcell (CP) ~\cite{Carr_Purcell} $\ $and the Carr--Purcell--Meiboom--Gill
(CPMG)~\cite{Meiboom_Gill} and it has stimulated the development of new ones
\cite{DD_1999, UDD_2007}. The simplest dynamical decoupling (DD) sequences
consist on the application of $\pi$ pulses on the system to revert the decay
due to the system-environment interactions. Many works have been done in order
to compare the performance of the CPMG-like sequences and de Uhrig (UDD)
sequences \cite{UDD_2007} when the system is coupled to a spin bath
\cite{cory_diamond_2010,alvarezpra2010}. Yang, Wang and Liu have made a
comprehensive theoretical and experimental description of the DD techniques
and its variations~\cite{yang_preserving_2010}.\textbf{ }

In this work we will make a critical review of recent attempts to extend the
decoherence time of spins in solid state NMR or EPR. We show that the role of
line inhomogeneity should not be underestimated to observe coherences reliably.

We summarize our own findings on anomalous long lived echo signals
\cite{barrett,barrettPRL07,barrettPRB07,belpat,belpat2}. In particular we
emphasize the use of a simple stimulated echo sequence to discern if one is in
the presence of true long decoherence times or if the coherences are being saved
as polarization and then recovered. This understanding allows us to develop a
new strategy for their use that is reported here. It makes use of magnetic
field gradients that enable to store part of the phase information as
polarization. We demonstrate that inhomogeneous samples in presence of
homogeneous fields behave in the same manner than homogeneous samples in
presence of inhomogeneous fields.

Additionally, the correct phase cycle to use in multipulse sequences in order
to measure the true T$_{2}$ in inhomogeneous samples is shown. This sequence,
whose result has been empirically known \cite{mlev4,conradi90,CPMG4}, should
be taken as a standard to obtain a clean T$_{2}$ measurement. Here, we explain
how the stimulated echoes are canceled out eliminating their main effects: the
long tails.

\section{\textbf{Basic NMR concepts}}

A typical NMR experiment requires at least two magnetic fields: a static field
B$_{0}$, conventionally in the z direction, and a perpendicular radiofrequency
(RF) field $\text{B}_{1}$ applied to induce transitions. The time t$_{p}$
during which the RF is on is defined as the pulse time and the tilting angle
is $\theta_{p}=\gamma\text{B}_{1}\text{t}_{p}$, where $\gamma$ is the
gyromagnetic factor. The eigenstates of angular momentum for a spin $1/2$
along the static field direction are $\left\vert \uparrow\right\rangle $ and
$\left\vert \downarrow\right\rangle $. Once the thermal equilibrium is reached
the population of state $\left\vert \uparrow\right\rangle $ is larger than
that of $\left\vert \downarrow\right\rangle $ indicating a net polarization of
spins along the static field direction. This thermal polarization will
determine the maximum magnetization that can be created in the system. Any
other situation in which the macroscopic magnetization points in an arbitrary
direction will eventually have a z-magnetization which reflects a non
equilibrium polarization. A particular but very relevant case is the
application of a $\frac{\pi}{2}$ or saturation pulse after which the
magnetization lies in the xy plane. The population of the states is
equilibrated and converted in a superposition state (a coherence) where the
eigenstates $\left\vert \uparrow\right\rangle $ and $\left\vert \downarrow
\right\rangle $ have the same probability and a well defined phase relation.
The essential difference between polarization and coherences is that, in a
very good approximation, polarization does not evolve under interaction with
the static field while the transverse magnetization (or coherence) does.

Thus, after a $\frac{\pi}{2}$ pulse is applied, transverse magnetizations
associated with different volume elements (called isochromats) evolve under
different Hamiltonians due to small inherent inhomogeneities and spin--spin
interactions. This evolution causes a dephasing among the spins in the sample
and a free induction decay (FID) with characteristic time T$_{2}^{\ast}$ is
observed. The dephasing caused by inhomogeneities is reversible, so that
refocusing is possible by applying appropriate pulse sequences such as the
Hahn echo~\cite{Hahn} [HE: ($\frac{\pi}{2}-\tau-\pi-\tau-$ echo)]. During the
evolution time $\tau$ a dephasing occurs and then a $\pi$ pulse is applied.
The net effect of this pulse is to change the sign of all the Hamiltonians
linear in spin operators, thus reversing the evolution of the isochromats
which refocus after the same time $\tau$. As a result a magnetization echo is
observed, i.e. the coherent state created after the initial $\frac{\pi}{2}$ pulse is
partially recovered after a time 2$\tau$. The amplitude of the echo obtained
after a single refocusing pulse is attenuated by homonuclear interactions represented by
 non linear terms in the system Hamiltonian. In solids the most important contribution
will be the spin--spin dipolar interaction. If the separation time $\tau$
between the pulses is increased a decay in the magnetization is observed with
a characteristic time T$_{2}$. This quantity is known as the spin--spin decay
time, transverse relaxation or decoherence time. Actually, T$_{2}$ represents
the decoherence of a single spin caused by its interactions with other spins
after canceling out the interactions linear in spins. Indeed, the spin--spin
interaction can also be reversed by suitable sequences. They extend the
decoherence time to a new timescale $\mathrm{T}_{3}>\mathrm{T}_{2}%
$~characterizing the ability to control the many-body dynamics
\cite{ZME92,LUP98,mrev,CORY48}. Since, these sequences revert many-body
interactions in recent years the quantum information community refers to them
as Loschmidt Echoes \cite{LUP98,PLURH00,JalPa,Seligman,Jacquod}. In any case,
the decoherence time T$_{2}$ is the characteristic time that should be very
well known before any attempt of quantum processing. It imposes a first
natural limit to quantum control, beyond which more sophisticated experiments
that restore phase information (e.g. decoupling and Loschmidt Echo time
reversal strategies) are needed \cite{ZME92,LUP98,mrev,CORY48}. Since 1950,
several sequences have been developed in NMR to obtain the T$_{2}$ decay time.
As the Hahn echo experiment is very time consuming, multipulse sequences that
enable the acquisition of the full decay with\ a single train of pulses such
as the Carr--Purcell~\cite{Carr_Purcell} $\mathrm{CP}:(\frac{\pi}{2})_{x}%
-[\tau-\pi_{x}-\tau-echo]_{n}$ and the
Carr--Purcell--Meiboom--Gill~\cite{Meiboom_Gill} $\mathrm{CPMG}:\left(  \frac
{\pi}{2}\right)  _{x}-\left[  \tau-\left(  \pi\right)  _{y}-\tau-echo\right]
_{n}$ are the most commonly used.
\section{\textbf{Long lived Signals}}

The interest in nuclear spins for QIP, in particular silicon~\cite{Kane},
focused on systems where nuclei couple to electron spins which are good
candidates for hybrid technologies. This motivated many NMR works where
typical experiments to measure T$_{2}$ were
reported~\cite{barrett,Watanabe,Ladd05}. In these works after applying the
CPMG like sequences it was found that it is possible to detect the spin-1/2
NMR signal of $^{29}$Si up to times much longer (more than two orders of
magnitude) than the characteristic decay time observed with the Hahn echo
sequence ($\approx5.6$ ms). This striking and promising finding came as a
surprise and triggered the realization of many experiments to understand its
origin. From the beginning it was verified that there is no dependence on the
amount of donors or acceptors in the $^{29}$Si polycrystalline sample even
when these could change the linewidths and spin-lattice relaxation time by one
order of magnitude. Experiments\ were repeated on $^{29}$Si single crystals
and Pyrex, as well as $^{13}$C in C$_{60}$ yielding the same behavior.

Since the measurements were performed in natural abundance $^{29}$Si (4.67\%)
and $^{13}$C (1.1\%), it was clear that the spins are diluted yielding weakly
coupled networks. Notice that the spin dipolar interactions decay with the
cube of the distance and that the gyromagnetic factors of $^{29}$Si and
$^{13}$C are approximately one fifth and one fourth of the $^{1}$H rendering
$^{29}$Si--$^{29}$Si and $^{13}$C--$^{13}$C couplings which are 1/25 and 1/16 of
the $^{1}$H--$^{1}$H respectively. This suggests that the disorder of the
chemical shift energies (or site energies) in such weakly coupled networks
could play a relevant role in the observed phenomena, which have not been
observed in the popular proton samples (ie. in solid state NMR of $^{1}$H the
usual observation is that the results from a Hahn echo coincides with those of
the CP or CPMG sequences). With this in mind we started a series of
experiments and calculations in a natural abundance polycrystalline sample of
C$_{60}$. The first experiment was a traditional Hahn echo measurement as a
function of the time $\tau$ which yielded T$_{2}^{HE}\simeq$ 15 ms. It should
be noted that the decay of the FID revealed a T$_{2}^{\ast}\simeq$ 2 ms,
indicating a high degree of inhomogeneties or reversible single spin
interactions. Then, we used a series of different trains, derived from the CP
and CPMG sequences These sequences were named respecting their origin (CP or
CPMG for $\pi$ pulses with $0^{\circ}$ or $90^{\circ}$ phase shifts with
respect to the first $\frac{\pi}{2}$ pulse, respectively) and considering the number
of phases ($1$ or $2)$ for the $\pi$ pulses in each cycle as follows:%

\begin{subequations}
\begin{align}
\mathrm{CP1}  &  :\left(  \frac{\pi}{2}\right)  _{x}-\left[  \tau-\left(
\pi\right)  _{x}-\tau-\text{echo}\right]  _{n},\label{CP1}\\
\mathrm{CP2}  &  :\left(  \frac{\pi}{2}\right)  _{x}-\left[
\begin{array}
[c]{c}%
\tau-\left(  \pi\right)  _{x}-\tau\ -\text{echo}-\\
\tau-\left(  \pi\right)  _{-x}-\tau-\text{echo}%
\end{array}
\right]  _{n},\label{CP2}\\
\mathrm{CPMG1}  &  :\left(  \frac{\pi}{2}\right)  _{x}-\left[  \tau-\left(
\pi\right)  _{y}-\tau-\text{echo}\right]  _{n},\label{CPMG1}\\
\mathrm{CPMG2}  &  :\left(  \frac{\pi}{2}\right)  _{x}-\left[
\begin{array}
[c]{c}%
\tau-\left(  \pi\right)  _{y}-\tau\ -\text{echo}-\\
\tau-\left(  \pi\right)  _{-y}-\tau-\text{echo}%
\end{array}
\right]  _{n}.\text{ } \label{CPMG2}%
\end{align}

Note that CP1 and CPMG1 are the original CP \cite{Carr_Purcell} and CPMG
\cite{Meiboom_Gill} sequences respectively.

We found that by applying the CPMG1 sequence or the CP2 sequence, a
magnetization tail appears, i.e. the signal remains for times much longer than
$T_{2}^{HE}$ as observed in $^{29}$Si, \cite{barrett}. Figure
\ref{fig1sinising} shows the normalized magnetizations acquired by applying
the sequences HE; CP1; CP2; CPMG1 and CPMG2 to polycrystalline C$_{60}$. In
the last four sequences $\tau=1$ms and signal acquisition is performed at the
top of the echo. In all the sequences the durations of the $\pi$ and $\pi
/2$\ pulses\ were carefully set from nutation experiments to 7.0 $\mu s$ and
3.4 $\mu s,$ respectively.

On the other hand, by applying the CP1 or the CPMG2 sequences the decay time
is not longer than $T_{2}^{HE}$, but it is remarkable that at short times the
magnetization makes a zigzag (consequence of the presence of stimulated
echoes) see inset in Fig.\ref{fig1sinising}. Besides, the magnetization
oscillates going through negative values before reaching its final asymptotic
zero value but keeping the $T_{2}^{HE}$ decay as envelope (see
Fig.\ref{oscilaenvolT2}).

\begin{figure}[ptb]
\centering
\includegraphics[width=9cm]{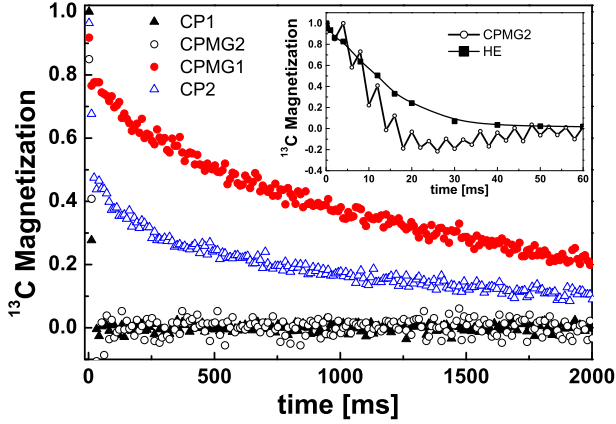}\caption{ Comparison of the
$^{13}$C magnetizations acquired with different $T_{2}\ $sequences ($\tau=1$
ms). The insert shows the decay observed with the Hahn echo and the short time
regime of the CPMG2 sequence.}%
\label{fig1sinising}
\end{figure}

By plotting these oscillating decays as a function of the echo number a match
of the frequencies can be observed, i.e. the maxima and minima overlap for the
same number of $\pi$ pulses (echo number, see Fig.\ref{oscila_T2_echonumber}).
This behavior reveals an accumulation of pulse errors which complete a cycle
after approximately 80 pulses. The amplitudes decrease with increasing $\tau$
following the behavior of the long tails. Note in Fig.\ref{fig1sinising} that
for $\tau=1$ $\mathrm{ms,}$the oscillation almost disapears.

\begin{figure}[ptb]
\centering
\includegraphics[width=9cm]{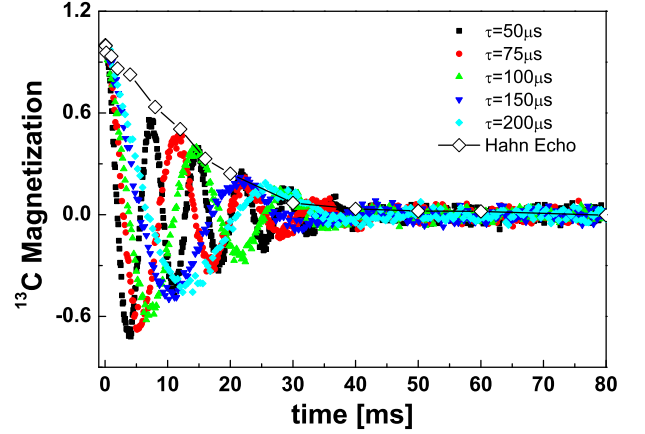}\caption{$^{13}$C Magnetization
decay as a function of time measured for different time windows with the CPMG2
sequence in polycrystalline C$_{60}$. It is evident that the envelope of the
oscillatory decays are dominated by T$_{2HE}$.}%
\label{oscilaenvolT2}
\end{figure}

\begin{figure}[ptb]
\centering
\includegraphics[width=9cm]{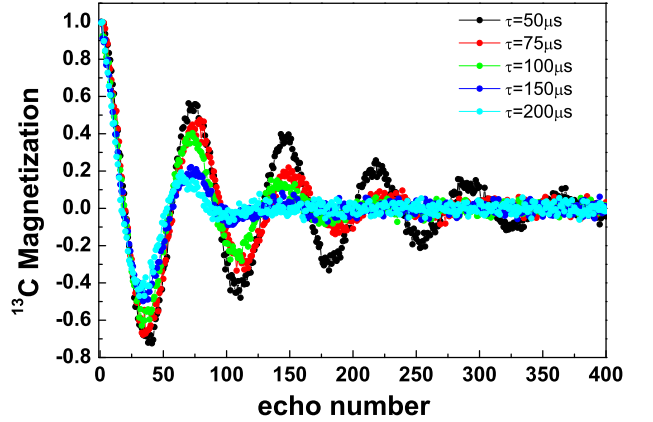}\caption{$^{13}$C
Magnetization decays as a function of the number of pulses (or echoes) for
short $\tau^{\prime}s$ in the CPMG2 sequence. }%
\label{oscila_T2_echonumber}
\end{figure}

Another unexpected observation was the presence of noticeable stimulated
echoes $[$STE$:\left(  \frac{\pi}{2}\right)  _{x}-\tau-\left(  \frac{\pi}%
{2}\right)  _{y}-t_{1}-\left(  \frac{\pi}{2}\right)  _{y}-t]$ when applying
pulse sequences of the form, $\ [\left(  \frac{\pi}{2}\right)  _{x}%
-\tau-\left(  \pi\right)  _{\varphi_{2}}-t_{1}\ -\left(  \pi\right)
_{\varphi_{3}}-acq]$ with variable $\tau$ and $t_{1}$. Indeed, these sequences
with $\pi$ pulses should not produce stimulated echoes. To remember the
essentials of the argument we reproduce here our results after the application
of the STE sequence with the phases of the CPMG1 and CPMG2, called CPMG1$^{STE}$
and CPMG2$^{STE}$ respectively in C$_{60}$ setting $\tau=1$ $\mathrm{ms}$ and
$t_{1}=15$ $\mathrm{ms}$ (see Fig.\ref{SEconsdestruc}). It can be observed that the
stimulated echoes appear $1$ $\mathrm{ms}$ after the last pulse (remember
that they should not be formed with perfect $\pi$ pulses) and they show a
phase that coincides or it is opposite to that of the Hahn echo, that appears
at $t_{1}-\tau=14$ $\mathrm{ms,}$ for the CPMG1$^{STE}$ and CPMG2$^{STE}$
respectively. A phenomenological explanation for the long magnetization tails
based on the constructive or destructive interferences between the stimulated
and\ \emph{normal }echoes was reported by us in a previous paper
~\cite{belpat} and later on verified through hole burning experiments
\cite{belpat2}. Thus, although dissapointing for the experimentalists of the
quantum information community, we concluded that the observed long tails in
the magnetization decays are not a signature of long decoherence times but of
the recovery of coherences saved as polarization. These long pseudocoherent
tails appear as a consequence of the formation of stimulated echoes. These
echoes, which need at least three pulses to appear, were first observed by E.
Hahn~\cite{Hahn} and have the particularity that can survive for times as long
as the spin-lattice relaxation time T$_{1}$. They will be revised in the next
section in order to try to save and recover coherences in a controlled way.

\begin{figure}[ptb]
\centering
\includegraphics[width=9cm]{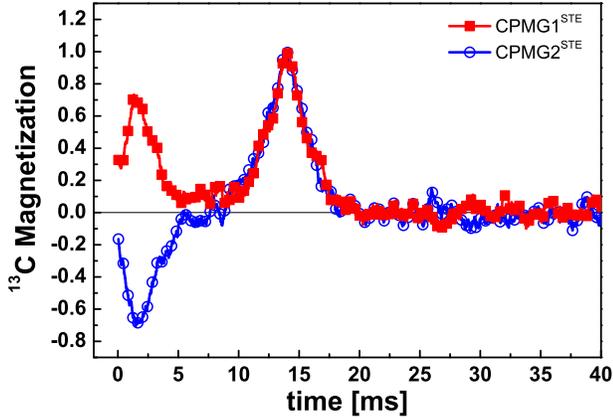}\caption{Stimulated echoes
observed with the CPMG1$^{STE}$ and CPMG2$^{STE}$ sequences in a C$_{60}$
polycrystalline sample. The parameters are $\tau=1$ ms and $t_{1}=15$ ms,
giving rise to a stimulated and a normal echo at $1$ ms and $14$ ms after the
last pulse, respectively. Notice the inversion of the phase in the stimulated
echo.}%
\label{SEconsdestruc}
\end{figure}

This is the key experiment, and we suggest that this stimulated echo sequence
with $\pi$ pulses, $[\frac{\pi}{2}-\tau-\pi-2\tau-\pi-\tau]$ should be used in
the same conditions of the CP-like train sequences to be able to discard the
stimulated echoes as the origin of the long magnetization tails. Although the
presence of the stimulated echoes after this sequence is not expected, they
can be generated by the pulse angle distribution present in an inhomogeneously
broadened line. In all the samples mentioned above, the spin--spin decay time
measured with a Hahn echo sequence (T$_{2\text{HE}}$)~\cite{Hahn} was about an
order of magnitude longer than the FID characteristic time (T$_{2}^{\ast}$),
evidencing the line inhomogeneity.

Magnetization tails have been observed in $^{29}$Si, C$_{60}$ and Y$_{2}%
$O$_{3}$~\cite{Ladd05,barrettPRL07,barrettPRB07}, in adamantane under $^{1}$H
decoupling \cite{gonzalo11} and indirectly in $^{31}$P in EPR
experiments~\cite{morton_solid-state_2008}.

The requirements for the formation\ of the long tails can be summarized as
follows: (a) an rf field inhomogeneity or a naturally inhomogeneous line able
to produce different tilting angles in different sites of the sample and (b)
the absence of or a very slow spin diffusion (noneffective flip-flop
interactions). Under these conditions the different tilting angles do not
average to the nominal $\pi$ pulses preset in the sequences leaving a
perpendicular component that behaves as in the regular stimulated echo
sequence $\left[  STE:\left(  \frac{\pi}{2}\right)  _{x}-\tau-\left(  \frac
{\pi}{2}\right)  _{y}-2\tau\ -\left(  \frac{\pi}{2}\right)  _{y}-acq\right]  $
as schematized in Fig.\ref{SE-spheres}. Still, one question that remained open
was if the relation between the spin-site energy difference and the dipolar
couplings (the main source of flip-flop interactions in solid systems) should
be verified in a local (microscopic) basis or if a smoothly varying
inhomogeneity that only satisfy the relation at longer distances could produce
the effect. To investigate this issue we resorted to the use of gradients in
homogeneous samples.

\begin{figure}[ptb]
\centering
\includegraphics[width=9cm]{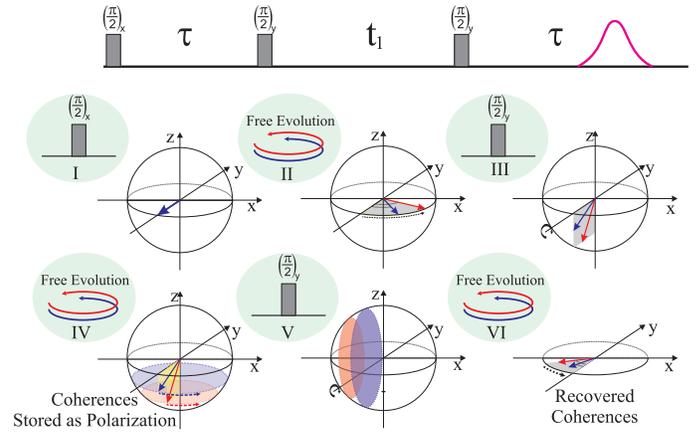}\caption{Traditional stimulated
echo sequence. The coherence states created during the first evolution period
$\tau,$ are stored as $z$-amplitude (polarization) during t$_{1}$and then
reconverted to coherences (transverse magnetization) after the third pulse.}%
\label{SE-spheres}
\end{figure}

{We artificially create the necessary conditions to observe the long tails
after a CPMG measurement, even for homogeneously broadened samples. The method
uses the fact that homogeneous lines, evolving under the influence of a field
gradient, are inhomogeneously broadened giving rise to stimulated echoes with
the consequent long tails. This method provides a controlled way to move from
homogeneous lines (for which the decay observed after a Hahn echo and CPMG
like sequences are in complete agreement) to inhomogeneous ones. We reach
inhomogeneous limits analogous to $^{29}$Si or C$_{60}$ cases, where we show
that careful interpretation of the experiments must be done before assigning
\textquotedblleft the true\textquotedblright\ spin--spin decay time.}

\section{Behaviour of the magnetization during a simple stimulated echo
sequence}

The stimulated echo contribution to the CPMG sequence has been widely studied
in presence of inhomogeneous B$_{0}$ and B$_{1}$ fields. It has been
demonstrated that due to the inaccuracy of the pulses and the off resonance
effects, the CPMG echoes have contributions not only from the Hahn echo but
also from signals similar to stimulated echoes arising from different
coherence pathways~\cite{HurlimannJMR00,HurlimannJMR01,SongJMR02}. This is in
complete agreement with our findings.

Here, we review in detail the magnetization evolution during a stimulated echo
to make evident how it can be potentially exploited to create memories.

The eigenstates of angular momentum for a spin $1/2$ along the static field
direction are $\left\vert \uparrow\right\rangle $ and $\left\vert
\downarrow\right\rangle $. Once the thermal equilibrium is reached the
population of state $\left\vert \uparrow\right\rangle $ is larger than that of
$\left\vert \downarrow\right\rangle $ indicating a net polarization of spins
along the static field direction. This thermal polarization will determine the
maximum magnetization that can be created in the system. Any other situation
in which the macroscopic magnetization points in an arbitrary direction will
eventually have a z-magnetization which reflects a non equilibrium
polarization. A particular but very relevant case is the application of a
$\frac{\pi}{2}$ or saturation pulse after which the magnetization lies in the
xy plane. The population of the states is equilibrated and converted in a
superposition state (a coherence) where the eigenstates $\left\vert
\uparrow\right\rangle $ and $\left\vert \downarrow\right\rangle $ have the
same probability with well defined phase relation. The essential difference
between polarization and coherences is that, in a very good approximation,
polarization does not evolve under interaction with the static field while the
transverse magnetization (or coherence) does.

{It is important to correctly understand the processes behind the stimulated
echo in order to separate its contribution from the Hahn echo contribution.}
To clarify the details, let us go back to the simplest stimulated echo sequence:%

\end{subequations}
\begin{equation}
\text{STE}:\left(  \frac{\pi}{2}\right)  _{x}-\tau-\left(  \frac{\pi}%
{2}\right)  _{y}-t_{1}-\left(  \frac{\pi}{2}\right)  _{y}-t
\end{equation}

In Fig.~\ref{SE-spheres} the stimulated echo formation is schematized using a
vectorial spin model. The model provides a pictorial view where it can be
observed that the magnetization is indeed stored as polarization and then
reconverted to transverse magnetization. The information on the coherence
state (characterized by a well defined phase relationship between the
transverse spins) created in the first evolution period, $\tau$, is conserved
during $t_{1}$ and refocused {as an echo during} the third period at $t=\tau$.

Let us consider a Hamiltonian linear in the spin operators with no
interactions between different spins,%

\begin{equation}
\mathcal{H}=\sum_{j=1}^{N}{\delta\omega_{j} I_{j}^{z}}. \label{hamiltonian}%
\end{equation}

If the delta pulse approximation is taken and we consider the system in
thermal equilibrium before the first pulse, the evolution of the density
matrix can be written as:%

\begin{equation}
\rho(\tau^{-})=-\sum_{j=1}^{N}\left[  \cos(\delta\omega_{j}\tau)I_{j}^{y}%
-\sin(\delta\omega_{j}\tau)I_{j}^{x}\right]  \label{coherence}%
\end{equation}

The expression above represents a coherent state (transverse magnetization),
which immediately after the second pulse becomes,%

\begin{equation}
\rho(\tau^{+})=-\sum_{j=1}^{N}{\left[  \cos(\delta\omega_{j}\tau)I_{j}%
^{y}+\sin(\delta\omega_{j}\tau)I_{j}^{z}\right]  }%
\end{equation}

After the second evolution period (t$_{1}$), the new state of the system can
be expressed as:%

\begin{align}
\rho(\tau+\text{t}_{1})=  &  -\sum_{j=1}^{N}\left\{  \cos(\delta\omega_{j}%
\tau)[I_{j}^{y}\cos(\delta\omega_{j}t_{1})-I_{j}^{x}\sin(\delta\omega_{j}%
t_{1})]\right. \nonumber\\
&  +\left.  {\sin(\delta\omega_{j}\tau)I_{j}^{z}}\right\}
\label{polarization}%
\end{align}

During the time t$_{1}$, there is a component of the initial magnetization
stored as non equilibrium polarization in the ${I_{j}^{z}}$. This component
does not evolve during t$_{1}$ under the chemical shift Hamiltonian in
Eq.~\eqref{hamiltonian}. The last term within the sum in
Eq.\eqref{polarization} defines a polarization grating which lodges in its
amplitude the information of the coherent state, i.e. the phase, of
Eq.~\eqref{coherence}. The useful interpretation is that the coherent state
created during the first evolution period remains frozen as a {local}
polarization amplitude during a long period $t_{1} $. It should be noted that
during the second period the polarization in Eq.~\eqref{polarization} is not
in thermal equilibrium (there is not a Boltzmann distribution because of the
$\delta\omega$ dependence). The spin-lattice relaxation, characterized by
T$_{1}$, will lead to a decay of the polarization grating during $t_{1}$.

The third $\left(  \frac{\pi}{2}\right)  _{y}$ pulse will convert the
polarization of Eq.~\eqref{polarization} in transverse magnetization as%
\begin{align}
\rho(\tau+\text{t}_{1}^{\text{+}})= &  -\sum_{j=1}^{N}\left\{  \cos
(\delta\omega_{j}\tau)[I_{j}^{y}\cos(\delta\omega_{j}t_{1})+I_{j}^{z}%
\sin(\delta\omega_{j}t_{1})]\right.  \nonumber\\
&  +\left.  {\sin(\delta\omega_{j}\tau)I_{j}^{x}}\right\}
\end{align}
which will evolve under the linear Hamiltonian influence as:%
\begin{align}
&  \rho(t+\tau+\text{t}_{1})=-\sum_{j=1}^{N}\cos(\delta\omega_{j}\tau)\left\{
\cos(\delta\omega_{j}t_{1})[I_{j}^{y}\cos(\delta\omega_{j}t)\right.
\nonumber\\
&  \left.  -I_{j}^{x}\sin(\delta\omega_{j}t)]+I_{j}^{z}\sin(\delta\omega
_{j}t_{1})\right\}  \nonumber\\
&  +{\sin(\delta\omega_{j}\tau)\left(  I_{j}^{x}\cos(\delta\omega_{j}%
t)+I_{j}^{y}\sin(\delta\omega_{j}t)\right)  }\label{stimecho}%
\end{align}

The last two terms in Eq.~\eqref{stimecho} do not depend on $t_{1}$ {and when
$t=\tau$ the phase differences acquired between the spins in the first
evolution period are refocused forming the stimulated echo.} As can be seen
from Eq.\eqref{stimecho}, if $t_{1}$=0 the sequence is reduced to a simple
$\left(  \frac{\pi}{2}\right)  _{x}-\tau-\left(  \pi\right)  _{y}-\tau-$ Hahn
echo sequence and all the initial polarization is again at the -y axis. On the
other hand, if $t_{1}\gg\text{T}_{2\text{HE}}$, but still shorter than the
spin-lattice relaxation time, only the term that do not depend on $t_{1}$
survives and the coherences created during the first evolution period are
refocused at $t=\tau$. Notice that if $2\tau\approx\text{T}_{2\text{HE}}$ the dipolar interaction can not be neglected and
the assumptions performed in our calculations, where only the chemical
shifts were considered, are not valid anymore.

It is straightforward to see that if the second and third pulses in
Fig.~\ref{SE-spheres} are $\pi$ pulses a stimulated echo will not be
created~\cite{mehring,belpat}. Nevertheless if a spatial distribution of flip
angles is present in the sample, stimulated echoes will be created due to the
deviations from $\pi$ pulses as shown in ref.~\cite{SongJMR02,belpat}. The
long tails previously observed in inhomogeneous samples are a consequence of a
pulse angle distribution among the sample. From Eq.\eqref{SE-spheres} it can
be deduced that the long tails observed are due to the formation of stimulated
echoes and its consequent storage of local polarization. For example, the
magnetization in C$_{60}$ was stored as polarization for times two orders of
magnitude longer than the decoherence time (T$_{2\text{HE}}$%
)~\cite{belpat,belpat2,barrettPRL07,barrettPRB07}.

\section{Controlled generation of pseudocoherences}

In this section, we introduce a method to artificially create, even in
homogeneous samples, the conditions under which the stimulated echoes are
formed. {The purpose of this method is to study in detail the amplitude of the
polarization stored and the time the polarization can be stored as a function
of the inhomogeneity added to the sample. Sequences can be developed in order
to take advantage of this stored polarization ~\cite{morton_solid-state_2008}}.

The system we have previously studied, C$_{60}$, presents an inhomogeneously
broadened line which satisfies $\text{T}_{2\text{HE}}\approx7T_{2}^{\ast}$.
Similar situations were observed in all the systems that manifest long decay
times when CPMG-like sequences are applied \cite{belpat,belpat2,barrett}.

Here, we use a homogeneous sample to mimic the behaviour observed in
inhomogeneous samples by adding magnetic field gradients. When a field
gradient, $G=\partial B_{z}/\partial z,$ is applied in the direction of the
static magnetic field B$_{0}$\textbf{z}, the z component of the magnetic field
is written as:%

\begin{equation}
B_{z}=B_{0}+zG. \label{Bzcongrad}%
\end{equation}

As a consequence of the field gradient presence, the spin frequency is
modified as follows:%

\begin{equation}
\omega(z)=\omega_{0}+\omega_{G}(z) \label{omegacongrad}%
\end{equation}

\noindent where $\omega_{G}(z)=\gamma Gz$, with $\gamma$ the gyromagnetic factor.

Similarly to Eq.\eqref{hamiltonian}, the field gradient produces a Hamiltonian
linear in spin operators. If a strong enough gradient is applied, the main
interaction for the spin evolution is:%

\begin{equation}
\mathcal{H}_{G}=\gamma\hbar G\sum{I_{i}^{z}z_{i}}. \label{hamgrad}%
\end{equation}

The presence of this Hamiltonian term causes an inhomogeneously broadened line
whose width depends on the applied field gradient.

{In the present work the sample used as example is polydimethylsiloxane
(PDMS)~\cite{acostaMacrom2009,acosta2010}. The characteristic time decays for
the PDMS are $T_{2}^{\ast}\approx0.6$~ms and $T_{2HE}\approx1.8$~ms indicating
a small inhomogeneity in the system. The condition~$\text{T}_{2\text{HE}%
}\approx7~\text{T}_{2}^{\ast}$ found in C}$_{{60}}${ is easily satisfied with
a field gradient of strength around 5 G/cm.} All the experiments were
performed at 7 Tesla in a Bruker AvanceII. A birdcage coil of 10 mm ID and
Bruker microimaging coils were used. The pulse durations were 12.5 $\mu$s for
the $\frac{\pi}{2}$ and 25.5 $\mu$s for the $\pi$ pulses.

\begin{figure}[ptb]
\centering
\includegraphics[width=9cm]{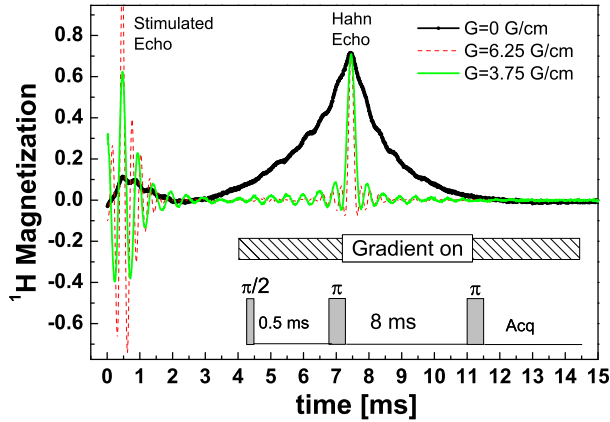}
\caption{Stimulated echo formation in presence of field gradients for the
three-pulse sequence shown at the bottom of the figure. Notice that while the
Hahn echo amplitude remains constant the stimulated echo increases with the
magnitude of the applied field gradient. }%
\label{seyhe}%
\end{figure}

To validate the proposed method we applied a $\left(  \frac{\pi}{2}\right)
_{x}-\tau-\left(  \pi\right)  _{y}-t_{1}-\left(  \pi\right)  _{y}$ sequence
with $\tau=0,5$~ms and $t_{1}=8$~ms with and without field gradient, see
Fig.\ref{seyhe}. The temporal windows are chosen such that the refocalization
of Hahn and stimulated echoes are produced at different times, remember that
for perfect $\pi$ pulses no stimulated echoes are expected. However, even in
absence of the external field gradient a small stimulated echo is observed due
to the sample inhomogeneity mentioned before. Notice that its amplitude is
approximately eight times smaller than the Hahn echo amplitude after
approximately 4$\times$T$_{\text{2HE}}$. Upon application of the external
field gradient $G$, the Hahn echo signal is spread into different frequency
components keeping the amplitude constant, while the stimulated echo increases
depending on $G$. It is clear that under the interpulse timings required by
the CPMG ($t_{1}$=2$\tau$) the STE overlaps the Hahn echo. By applying proper
phase cycling schemes, the discrimination of these two echoes becomes
possible~\cite{belpat2}.

\begin{figure}[ptb]
\centering
\includegraphics[width=9cm]{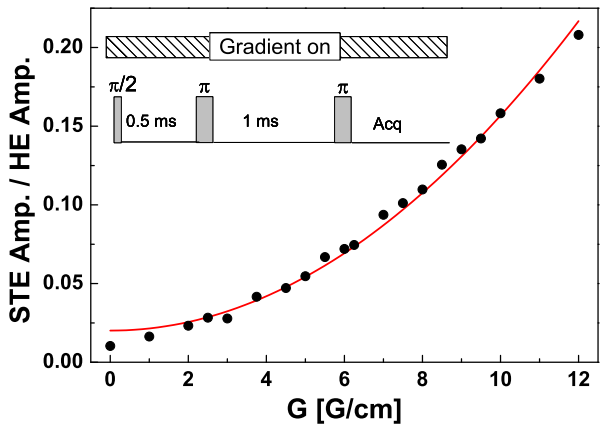}
\caption{Stimulated echo amplitude normalized with the corresponding
Hahn echo amplitude as a function of the applied field gradient. By
fitting the data with the function $a\text{G}^{2}+b$ the parameters
$a=0.00137\left(  \frac{\text{G}}{\text{cm}}\right)  ^{-2}$ and $b=0.02$ are
obtained, straight line in the plot.}%
\label{se}%
\end{figure}

In Fig. \ref{se} the STE is normalized with the amplitude of the HE to give
its relative contribution as a function of the applied gradient field $G$. Two
datasets were acquired with the proper phase cyclings in order to obtain the
stimulated and the Hahn echoes independently. It is noticeable that the
stimulated echo grows quadratically with the magnitude of the applied field gradient.

\section{Long tails generated using magnetic field gradients}

\begin{figure}[ptb]
\centering
\includegraphics[width=9cm]{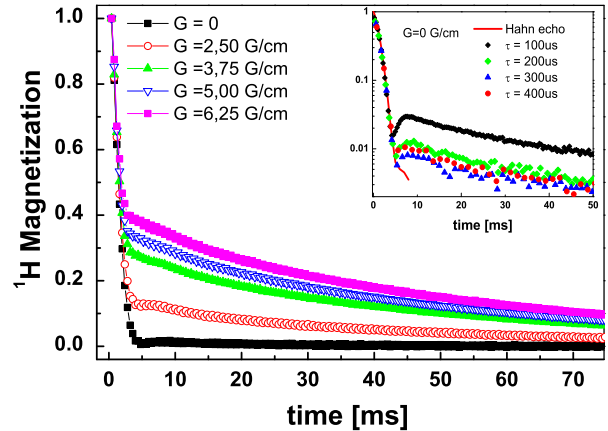}
\caption{Magnetization as a function of time for $\tau=200$
$\mu s$ and several field gradient strengths in the CP2 sequence. The Hahn echo and CP2 decays are very similar in
the absence of field gradients. The inset shows the evolution of the magnetization 
for different values of $\tau$ when $G=0$. Notice the change of the trend in the amplitudes between $\tau=300$ and $\tau=400$ $\mu s$ }%
\label{colas200u}%
\end{figure}

As reviewed in Section III, and Figs.1-4, we analyzed~\cite{belpat}, two
different sequences derived from Carr--Purcell~\cite{Carr_Purcell} and
Carr--Purcell--Meiboom--Gill~\cite{Meiboom_Gill} which produce long decay times
compared to the Hahn echo in inhomogeneous samples, namely CP2 and CPMG1
(see Fig.\ref{fig1sinising}). In this section, the evolution of the
magnetization during these sequences is studied upon variations of external
field gradients and interpulse spacing. In absence of an external field
gradient the decays observed with the multipulse CPMG1 and CP2 and the Hahn
echo sequences in PDMS are almost the same.
There is however a low contribution (c.a. 3$\%$) with decay time of 40 ms that 
can be observed by using a logarithmic scale in the inset of Fig.~\ref{colas200u}. This is a consequence 
of the very small contribution of the stimulated echo generated by the small inhomogeneity 
of the line even when $G=0$. Notice that the amplitudes of the tails decrease from $\tau=100$~$\mu s$ to $\tau=300$~$\mu s$, while at $\tau=400$~$\mu s$ this trend reverses.

In Fig.~\ref{colas200u} we show the dependence of the magnetization as a function of time for different field gradientes for the CP2 sequence at a fixed value $\tau=200$~$\mu s$.
As stronger field gradients are applied during the complete pulse
train for both CP2 and CPMG1, bigger amplitudes of the slow decaying
contributions appear manifesting the storage of coherences. 

As observed in C60, the magnetization decay in PDMS under gradients manifests
two different contributions. For short times, the Hahn echo is dominant while
for long times only the stimulated echo remains, storing polarization which
is then recovered as transverse magnetization. In order to characterize the
decays the data were fitted with a double exponential, where the short
characteristic time was fixed to the T$_{2HE}$ decay time.
\[
M(2\tau)=A_{s}\exp(2\tau/\mathrm{T}_{2HE})+A_{l}\exp(2\tau/\mathrm{t}_{l})
\]

\begin{figure}[ptb]
\centering
\includegraphics[width=9cm]{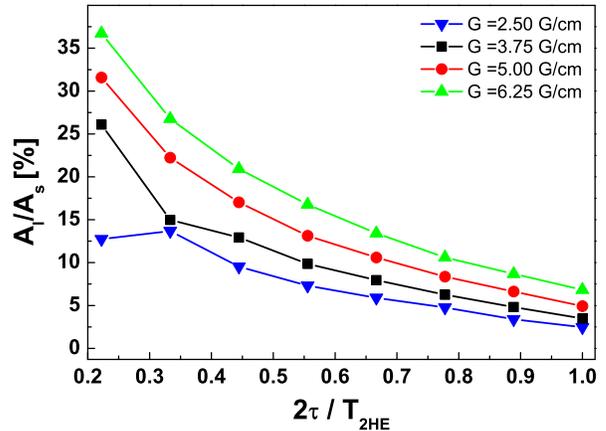} \caption{Percentage of magnetization
stored as polarization as function of the inter-pulse time 2$\tau$ scaled by
${\text{T}_{2HE}}$ in the multipulse sequence CP2. Better performance is
achieved by increasing the field gradients and/or decreasing $\tau$.}%
\label{fig3}%
\end{figure}

\begin{figure}[ptb]
\centering
\includegraphics[width=9cm]{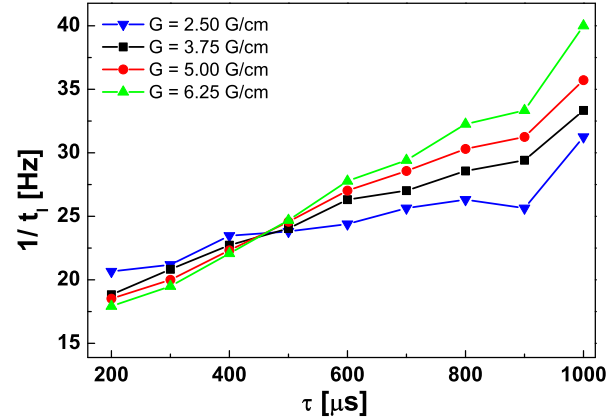}\caption{Characteristic linewidth
${1/\text{t}_{l}}$ as function of $\tau$ in the multipulse sequence CP2. Note
the crossing point at approximately 450 $\mu$s}%
\label{invtlvstau}%
\end{figure}

The parameters A$_{s}$, A$_{l}$ and t$_{l}$, where the subindexes \textit{s}
and \textit{l} stand for short and long decays respectively, were calculated
by fitting each dataset individually . The parameters A$_{l}$ and t$_{l}$ are
signatures of the stimulated echoes manifested through the long tails. A more
detailed analysis of the long tails was achieved by discarding the short time
points and fitting to a single exponential. In Figs.~\ref{fig3} \ and
\ref{invtlvstau} the percentage of magnetization stored as polarization
($A_{l}/A_{s}\times100$) and the corresponding values for the inverse of the
characteristic long decay times t$_{l}$ respectively are shown as a function
of the inter-pulse time scaled by T$_{2HE}$ in the CPMG2 sequence. From
Fig.\ref{fig3}, it can be seen that for the strongest gradient applied and for
the shortest $\tau$, more than $35\%$ of the initial magnetization was stored
as polarization and then refocused during the multipulse sequence. The
characteristic decay time t$_{l}$ of the long tails, is interpreted as a
storage time and the scaling with T$_{2HE}$ emphasizes the fact that the
magnetization can be preserved for times an order of magnitude longer than
the single-spin decoherence time. Under favorable experimental conditions
$2\tau/$T$_{2\text{HE}}$ can be decreased to get a better t$_{l}%
/$T$_{2\text{HE}}$ ratio. This amount can also be improved when shorter delays
$\tau$ are experimentally affordable.

\begin{figure}[ptb]
\centering
\includegraphics[width=9cm]{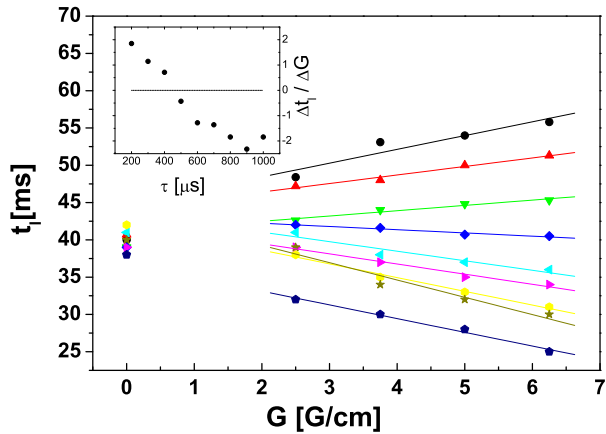}
\caption{Characteristic long decay
rate as a function of the field gradient. Different values of $\tau,$ which scale the
dipolar dynamics ($d_{\mathrm{eff.}}\simeq d^{2}\tau)$, from top to bottom,
200, 300, 400, 500, 600, 700, 800, 900 and 1000 $\mu\mathrm{s}$ are displayed.
Note the change of slope between 400 and 500 $\mu\mathrm{s.}$ This is
emphasized in the inset where the slopes are plotted as a function of $\tau$. The change of sign identifies the critical value. The decay of $t_l$= 40ms for $G=0$ is due to line
inhomegeneities and shimming imperfections}%
\label{slopevstau}%
\end{figure}

A remarkable behaviour of  1/t$_{l}$ vs. $\tau$ is
plotted Fig.\ref{invtlvstau} for different values of the gradient $G$. A change of regime occurs
at approximately $\tau_{c}=450$ $\mu s$ corresponding to t$_{l}\approx40$ ms,
the same relaxation time that is obtained in the absence of gradients. For values of
$\tau$ shorter than $\tau_{c}$, the linewidth $\Delta\nu$ decreases (i.e. the
characteristic long time t$_{l}$ increases) as the magnitude of the gradient
increases (see Fig. \ref{slopevstau}). For times longer than $\tau_{c}$, the
behavior reverses, i.e. t$_{l}$ decreases as $G$ increases. This is more clearly
appreciated in Fig.\ref{slopevstau} where the values of
\ t$_{l}$ for different values of $\tau$ are plotted vs. the magnitude of the
gradient $G$. It is seen that for $\tau\leq400\mu$s t$_{l}$ increases with the
gradient while for $\tau\geq500\mu$s t$_{l}$ decreases with $G$. This $\tau_{c}$
(interpulse time 900 $\mu$s) should be related to the characteristic
fluctuation time of the bath $\tau_{b}$, or its spectral density, essentially
dominated by dipolar interactions partially averaged by motion. In a simple
picture, one can interpret that a fast interpulse rate (1/(2$\tau_{c}$)$\gg
$1/$\tau_{b}$) produces a decoupling of the system from the bath, leading to
narrower lines (longer t$_{l}$ values) while the opposite condition leads to a
broadening of the line. In other terms, one can assume that at short values of
$\tau$ ($\tau\ll1/d)$ where $d$ stands for the magnitude of the dipolar
interaction, there is an \ "effective dipolar interaction" given by
$d_{eff}=d^{2}\tau$ as the pulses act as a breaking time or
1/$\tau$ as an exchange rate in an exchange narrowing process
\cite{Anderson54}. In this regime, dipolar spin dynamics are very slow. When
$\tau\geq1/d$, the spin diffusion, which is able to average the deviation of
the $\pi$ pulses which lead to the formation of stimulated echoes, is also
effective to blur out the gradients in the sample. Then, even with larger
gradients, the values of t$_{l}$ decrease spoiling the preservation of
coherences as polarization. While a deeper analysis of how the dynamical
variables $d$, $\tau$, and $\gamma G$ interact is beyond the scope of this
article, one conclusion is clear: in order to save coherences as polarization
the shortest $\tau$ and the biggest G should be used. This can even allow a
manipulation of nuclear spins with better spatial resolution. As it is clear
from Eq.\eqref{stimecho}, and schematized in Fig.\ref{SE-spheres}, the
STE sequence is not able to recover the whole coherence, but at most 1/2 of
its initial amplitude. Consequently, this strategy is not useful to make
 "quantum memories" (fidelity below 0.66). Nevertheless, it is potentially
useful to make read out memories. Moreover, in a different context, Kobayashi\textbf{
}et al. \cite{Kobayashi2009} have designed variations of the STE sequence that
overcome the 1/2 factor and recover a much bigger fraction of the coherences
(or transverse magnetization).

Finally, we address the possibility of performing experiments with multipulse
sequences that do not produce coherence storage as polarization. A four
phase cycle sequence that resembles the MLEV-4 \cite{mlev4}:
\begin{align}
\text{CPMG4}:\left(  \frac{\pi}{2}\right)  _{x}  &  -\left(  \tau-\pi_{y}%
-\tau-\text{echo}\right. \nonumber\\
&  -\tau-\pi_{y}-\tau-\text{echo}\nonumber\\
&  -\tau-\pi_{{-y}}-\tau-\text{echo}\nonumber\\
&  -\tau-\pi_{{-y}}-\tau-\text{echo}\left.  {}\right)  _{n} \label{CPMG4}%
\end{align}
has been empirically shown to be the most suitable for preventing unexpected
artifacts in a T$_{2}$ measurement~\cite{conradi90,CPMG4}.

\begin{figure}[ptb]
\centering
\includegraphics[width=9cm]{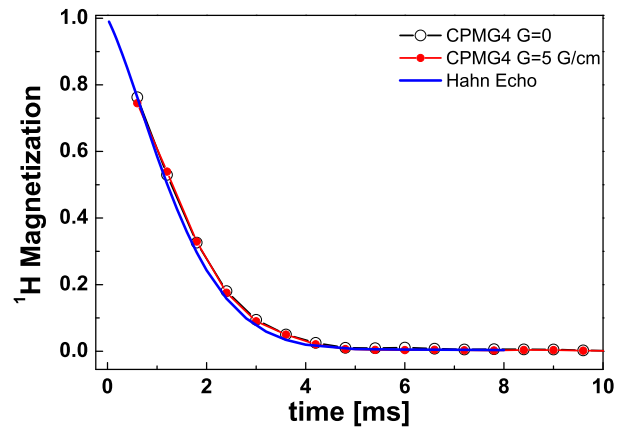} \caption{Magnetization observed by
applying the CPMG4 with gradient off and on compared to the Hahn Echo decay.}%
\label{fig4}%
\end{figure}

In Fig.~\ref{fig4} the results of a Hahn echo and CPMG4 measurements with and
without field gradient are plotted. It can be observed that the magnetization
decay agrees perfectly well for the three cases. The reason why this sequence
is the appropriate to measure T$_{2}$ can be understood from the stimulated
echo perspective presented previously~\cite{belpat}. The CPMG4 sequence does
not accumulate magnetization contributions from the constructive interference
between normal and stimulated echoes, and this is clearly manifested in the
results shown in Fig.~\ref{fig4}. Even in the presence of an inhomogeneous
field, long tails are not observed when this sequence is applied.

\section{Conclusions}

{In this article, we have reviewed the conditions for the formation of
stimulated echoes and the consequent generation of long tails in the magnetization decay. Besides, we report an original set of experiments using field gradients in homogeneous samples that extend and validates the previous findings and their interpretation. Moreover, they suggest a controlled way to preserve coherences as polarization that may account for more than 1/3 of the original signal. The long tails observed when multipulse sequences as CPMG are applied to
inhomogeneous samples were assigned to polarization that encodes quantum
coherences. In this way the coherences are stored for times much longer than
T$_{2\mathrm{HE}}$. It has been shown that the responsible of the long tails
are the stimulated echoes, formed due to the inhomogeneity of the sample. The requirements to observe these slow magnetization decays are (1) an rf field
inhomogeneity or a highly inhomogeneous line able to produce different tilting
angles in different sites of the sample and (2) non-effective flip-flop
interactions. Under these conditions the differences in resonance frequencies
are larger than the dipolar couplings, making ineffective the flip-flop
mechanism. Consequently, the different deviations from $\pi$ pulses
(regardless of its origin) cannot be averaged out by dipolar
interactions. For $2\tau\approx\text{T}_{2\text{HE}}$ the amplitude of the
coherence stored is less than $5\%$. In this condition,the time between pulses
is enough for the dipolar interaction to be operative. }

 In this paper we showed that the stimulated echoes preserve coherent states frozen in
the direction of the static magnetic field, then bring them back to the plane
where they are refocused and observed. It is important to remark that the
magnetization decay is very long because the coherences are stored as
polarization. So, these long decays should not be interpreted as a long
decoherence time during which quantum operations can be performed. However,
this strategy can be
exploited to design memories in nuclear or electron spin
systems. Then, upon recovery of the stored coherences, interactions
among the whole system can be turned on. The concept introduced in this paper
was indeed indirectly exploited in Ref.~\cite{morton_solid-state_2008}. There,
coherences induced in electron spins were transferred to a coherent state
between nuclear spins, stored by applying the CPMG1 sequence and transferred
back to electron spins.

The method that mimic the conditions to induce the stimulated echoes for
homogeneous samples in a controlled way is based on the application of field
gradients that produce a frequency distribution across the sample. This, in
turn, generates a pulse angle distribution that sets the conditions under
which the stimulated echoes appear. An important fact deduced from these
experiments is that the origin of the inhomogeneity of the samples is not
relevant for the observation of long tails. Whether the inhomogeneity arises
from a random distribution of nuclear spins (as occurs in $^{29}$Si) or on a
linear dependent gradient, like the one used in the present experiments, the
consequences are the same. Moreover, the use of suitable field gradients could
allow a more selective spatial manipulation of the memories. This method was applied to a PDMS sample,
for which we were able to perform a complete quantification of the procedure,
i.e. the storage time and ratio of coherence survival. There, it was possible
to detect a transition where the effective dipolar interaction $d^{2}\tau$
becomes of the order of the main frequencies in the bath's spectral density.
This imposes a limit in the interpulse spacing in which coherences can be
stored. Under the condition $\tau<1/d$ higher polarization storage and longer
decay times are achieved by using stronger gradients.

Finally, we tested the CPMG4 sequence under field gradients that reproduce inhomogeneous properties. The mechanism behind the
correct operation of this pulse sequence can be understood under the
stimulated echoes interferences. The
true single-spin T$_{2}$ decay time can be measured under this pulse sequence
either for homogeneous or inhomogeneous samples. Thus, we verified the empirically known robustness of this sequence which, regardless of the gradient, yields the same results as the Hahn echo. 

We acknowledge support from Fundaci\'{o}n Antorchas, CONICET, FoNCyT, MinCyT-Cor
SeCyT-UNC and the Partner Group for NMR Spectroscopy with High Spin
Polarization FaMAF-MPIP. We thank M. A. Villar, E. M. Vall\'{e}s and D. A.
Vega for sample preparation, A. D. Dente for helping in the design of Fig. 5,
 G. A. Monti and F.M. Pastawski for helpful discussions. \bigskip

\bibliographystyle{unsrt}

\end{document}